\documentstyle[twocolumn,prl,aps,psfig]{revtex}
\input{epsf}                                                               
\begin{document}
\draft

\title{\Large \bf A direct measurement of short range NN correlations in nuclei via
the reaction  $^{12}C(p,2p+n)$.}

\vspace{0.5cm}

\author{J. Aclander$^1$, J. Alster$^1$,
D. Barton$^2$, G. Bunce$^2$, A. Carroll$^2$,
 N. Christensen$^3$\thanks{Current address: Department of Physics, University of Auckland, New Zealand.},
H. Courant$^3$,  S. Durrant$^2$, S. Gushue$^2$, S. Heppelmann$^4$,
E. Kosonovsky$^1$, I. Mardor$^1$,Y. Mardor$^1$,  M. Marshak$^3$, Y. Makdisi$^2$, E.D. Minor$^4$\thanks{Current address: ``Concurrent Technologies Corporation'', Johnstown, PA, USA.}
,
I. Navon$^1$, H. Nicholson$^5$, E. Piasetzky$^1$, T. Roser$^2$, J. Russell$^6$,
C.S. Sutton$^5$, M. Tanaka$^2$\thanks{Deceased}, C. White$^3$, J-Y Wu$^4$\thanks{Current address: ``Fermi National Accelerator Laboratory''.}.}

\date{}
\maketitle

{\it $^1$School of Physics and Astronomy, Sackler Faculty of Exact
Sciences, Tel Aviv University.} {\it $^2$Brookhaven National
  Laboratory.} {\it $^3$Physics Department, University of Minnesota.} {\it  
$^4$Physics Department, Pennsylvania State University.} {\it $^5$Mount Holyoke
College.} {\it $^6$Physics Department, University of Massachusetts Dartmouth.}

\begin{abstract}
  
 The reaction $^{12}C(p,2p+n)$  was measured at beam momenta of 5.9
and 7.5 GeV/c.. We established 
the quasi-elastic character  of the reaction $C(p,2p)$  at $\theta_{cm}\simeq
90^o$, in a kinematically complete measurement. The neutron momentum was
measured in triple coincidence with the two emerging high momentum protons. 
We present the correlation between  the momenta of the struck target proton
and the neutron.  The events are associated with the high momentum components
of the nuclear wave function. We conclude that two-nucleon short range
correlations have been seen experimentally. The conclusion is based on
kinematical correlations and is  not based on specific theoretical models.

 \end{abstract}

\pacs{21.30.-x, 25.40.Hs, 24.50.+g}                                        

\narrowtext

Most nuclear properties can be described successfully within the single
particle model.  However, the simple picture breaks down when some detailed
features are studied, especially in the extreme regions of the nuclear wave
function. Experimental evidence for this comes from several sources. 
There is  additional strength at large nucleon momenta 
and large excitation energies 
\cite{kn:I0,kn:I10,kn:I101}. 
There are measurements of longitudinal to 
transverse response functions \cite{kn:I0}. 
All of these are inconsistent with the single particle model. 

 In recent years it has become clear that  two-nucleon correlations (not
necessarily of short range) play a role in nuclei.  Some examples of data
to that effect are the inclusive $(e,e')$ scattering at x$>$1 
\cite{kn:I5,kn:I51,kn:I52,kn:I53}, the semi-exclusive $(e,ep)$ measurements 
\cite{kn:I6,kn:I61,kn:I62,kn:I63}, real photon absorption
$(\gamma,2N)$ measurements
\cite{kn:I7,kn:I71} and the two-nucleon knockout reactions
$(e,ed)$, $(e,epp)$ and $(e,epn)$ 
\cite{kn:I8,kn:I81,kn:I82,kn:I83}. 
There is also evidence  from hadronic interactions such as pion absorption
\cite{kn:I9,kn:I91}, and backward scattered protons from a variety of
projectiles \cite{kn:I11}, including neutrinos \cite{kn:I12}. Since the early
theoretical work of Jastrow \cite{kn:I13}, there is a continuing theoretical
effort \cite{kn:I14,kn:I141,kn:I142,kn:I143,kn:I144,kn:I145} to relate the
experimental data to short range correlations (SRC) (especially two-nucleon
SRC) .

   The SRC between two nucleons in nuclei are a very elusive feature in
nuclear  physics. Their identification is very difficult because  they are
usually small compared to the single  particle components. It is also
difficult to separate SRC from effects such as meson exchange currents and
$\Delta$ components in the nucleus. In this paper we  describe how one can 
determine experimentally the existence of two-nucleon short range 
correlations (NN SRC). We present data that
show that they have been identified in the quasi-elastic A$(p,2p+n)$
reaction at large momentum transfer. 

 In a quasi-elastic (QE) scattering reaction with a high energy projectile,
 a single target nucleon with momentum $p_m$ is removed from the  
nucleus leaving
the residual nucleus at an  excitation energy $E_m$.  There are theoretical
expectations as well as experimental indications that NN SRC
 become important at large $p_m$ and $E_m$. 
Thus far the evidence for NN SRC is based on models that add SRC
contributions  to explain discrepancies with single particle interactions.
It is possible to obtain a direct experimental  signature for 
SRC by looking at the decay of the residual nucleus after the fast removal
of one of its nucleons by the  QE reaction. If  a high momentum target
nucleon is correlated with a partner nucleon, the partner will recoil with
a momentum {\bf $p$} in the direction  opposite 
to ${\bf p}_m$:  ${\bf p}_m=- {\bf p}$,
$p_m>k_F$, and $p>k_F$,  where $k_F$ is the Fermi momentum surface. 
Since $k_F \simeq  220$ MeV/c  (for $^{12}C$),  the signature will consist 
of two nucleons,
each with high momentum ($>220$ MeV/c) and a relative momentum of more
than 440 MeV/c,  balancing their momenta.

     We measured the high-momentum transfer quasi-elastic $C(p,2p)$ reaction
at $\theta_{cm}\simeq 90^o$ on carbon for 5.9 and 7.5 GeV/c incident protons,
in a kinematically complete coincidence experiment. The three-momentum 
components of both high $p_t$ final state protons were measured, which 
determined the missing energy and momentum of the target proton in the
nucleus. 
We measured the directions and momenta of neutrons in coincidence with these
two protons.
  The experiment (E850) was performed at the AGS accelerator at Brookhaven
National Laboratory with the EVA spectrometer 
 \cite{kn:ref3,kn:ref4,kn:ref5,kn:ref7},
which was located in the secondary beam line C1. The beam passed through a
sequence of two differential Cerenkov counters which identified the incident
particles. A scintillator in the beam served as  
timing reference. The beams ranged in intensity from 1 to 2$\times 10^7$ 
particles over a one second spill, every 3 seconds.  
The spectrometer consists of
a super-conducting solenoidal magnet operated at 0.8 Tesla. The beam enters 
along the symmetry axis of the magnet ($z$). 
The scattered particles are tracked by straw tube drift chambers which 
provide the transverse momenta of the particles and their scattering
angles. Details on  EVA spectrometer are given in Refs. 
\cite{kn:ref3,kn:ref4,kn:ref5,kn:ref7}.    Three
solid targets, $CH_2$, $C$ and $CD_2$ (enriched to $95\%$) were placed on
the $z$ axis,  separated by about 20 $cm$. They were
5.1$\times$5.1  $cm^2$  and 6.6 $cm$ long in the $z$ direction except for
the $CD_2$ target which was 4.9 $cm$ long. Their positions were interchanged
 several times at regular intervals. The data discussed here come from the
middle  and upstream target positions with $C$ and $CH_2$ only.
The downstream target geometry was not favorable for the correlation
 measurement.

Quasi-elastic scattering 
events , with only two charged particles in the
spectrometer, were selected. An excitation energy of the residual
nucleus $\mid E_{miss}\mid$$<$ 500 MeV  was imposed in order to suppress 
events where additional particles could be produced without being
detected in EVA. See Refs. \cite{kn:ref5,kn:I101} for details.

\begin{figure}[htbp]
\centerline{\epsfxsize=9cm  \epsfbox{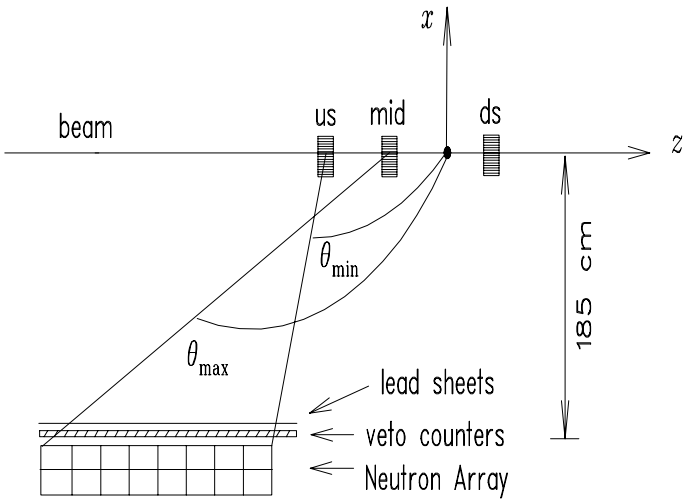} }
\vspace{1.0cm}
\caption
{
  The neutron counter  set up. The z axis is the symmetry axis of
the EVA spectrometer. The spectrometer itself
is not shown. $\theta_{min}$=102 $^o$ and $\theta_{max}$ 
= 125 $^o$. The three target positions are labeled as  upstream (us),
middle  (mid) and downstream (ds) positions.
}

\label{fig:1}
\end{figure}

 We measured the emerging neutrons in (triple) coincidence with the two
emerging high momentum protons. In  Fig. \ref{fig:1}, we present a
schematic picture of this setup. Below the targets 
 we placed a series of 16
scintillation bars covering an area of 0.8$\times$1.0 m$^2$ and 0.25 m deep. 
They
spanned an angular range of 102 to 125 degrees from the target. These
counters measured the neutron momenta by time of flight (TOF) with a TOF
resolution of $\sigma$= 0.5 nsec. This corresponds to a momentum
resolution of $\sigma$= 30 MeV/c at the highest momentum.
A set of veto counters served to eliminate charged particles. Lead sheets 
of 1.7 radiation lengths
were placed in front of the veto counters in order to reduce the number of 
photons entering the TOF spectrum.
A clearly identified peak at about 3 nsec per meter flight path,
due to remaining photons from the targets, was used for
calibration and to measure the timing resolution. We applied a cutoff in the 
TOF spectrum at 6 nsec per meter flight path, keeping neutrons below 600 MeV/c,
in order to eliminate any photons.

An example of a triple coincidence event which displays a NN SRC,
is shown in Fig. 2.
We show 
the transverse components ${{\bf p}_t(p_1)}$ and ${{\bf p}_t(p_2)}$ of the 
two outgoing high momentum protons as they were reconstructed in the 
trajectory analysis. The transverse momentum component of the struck 
target proton ${{\bf p}_t(p)}={{\bf p}_t(p_1)}+{{\bf p}_t(p_2)}$ 
and the component of the neutron on the plane perpendicular to the beam
(${\bf p}_t(n)$)
are drawn as well.

\begin{figure}[htbp]
\centerline{\epsfxsize=9.cm  \epsfbox{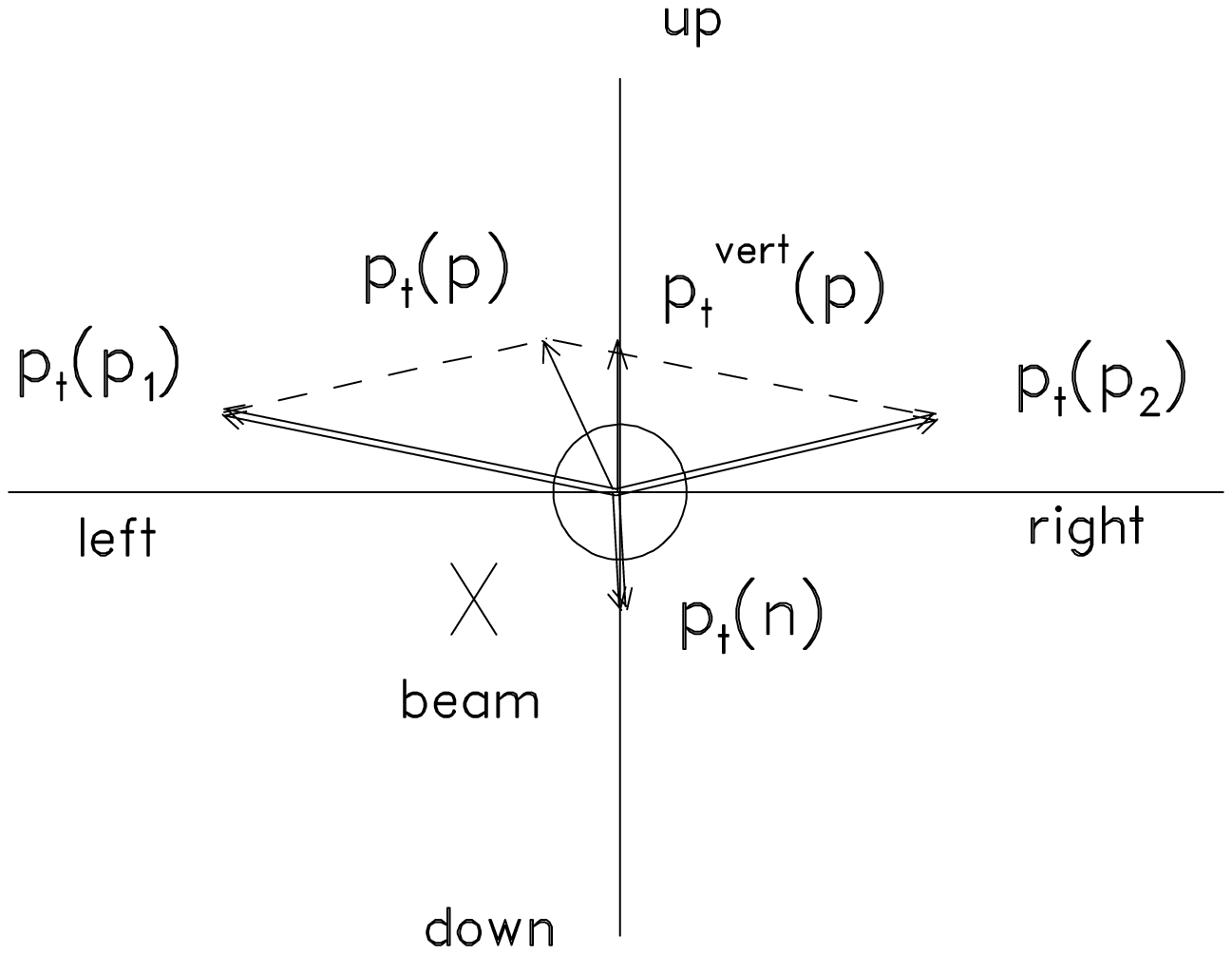} }
\vspace{1.0cm}
\caption{
 A triple coincidence event measured at an incident
momentum of 5.9 GeV/c . The vectors are projections on  the plane normal to
the incident beam. The axes are the vertical and horizontal
directions in that plane. 
The   ${{\bf p}_t(p_1)}, {{\bf p}_t(p_2)}$ are the transverse momenta of the
outgoing protons, ${{\bf p}_t(p)}$ is the transverse momentum  of the
target proton before the interaction and $p_t^{vert}(p)$ is its vertical
component. ${\bf p}_t(n)$ is the projection of the  neutron momentum on
the same plane. The circle indicates the scale for a momentum of 220 MeV/c.}
\label{fig:2}
\end{figure}

\begin{figure}[H]
\centerline{\epsfxsize=9.cm   \epsfbox{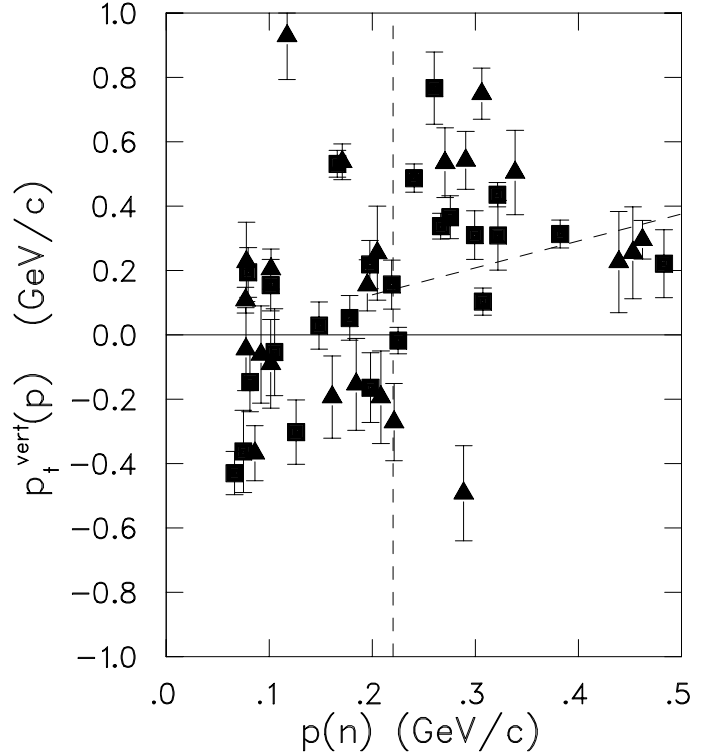} }
\vspace{0.5cm}
\caption{
The vertical component of the target nucleon  momentum  vs. the
total neutron momentum. The positive vertical axis is the upward direction.
 The events shown  are for triple coincidences of
the neutron with the two high energy protons emerging from the  QE $C(p,2p)$
reaction. The squares are for the  5.9 GeV/c incident beam  and the
triangles are for 7.5 GeV/c.  The dashed lines are explained in the text.
We associate the events in the   upper  right corner  with NN SRC.  }

\label{fig:3}
\end{figure}

 If the struck proton was correlated with a nearby neutron
and the $pn$ pair is at rest, the neutron
will emerge in the direction opposite to that of the struck proton and
with the same magnitude  ${\bf p}(n)$= - ${\bf p}(p)$. If the correlation
is of short range nature, we also expect  both   $p$(n) and  $p$(p) to be
above the Fermi sea level.  
Most of the events we measured do not have these ideal characteristics.
The angular correlation is spread out due to limited experimental
resolution, to center of mass motion of the $pn$ pair in the nucleus and to
final state interactions (FSI) of the outgoing protons.  Notwithstanding
the  inevitable smearing of the angular correlation, we can extract
information on the $np$ correlation from our data  by relaxing somewhat the
stringent "back to back" requirement. None of the above effects is
sufficiently large to prevent us from determining whether the momentum of
the target proton pointed upwards or downwards.  All the neutrons are
detected in the downward direction. Consequently,  we concentrate on the
vertical (up-down) component of the struck proton. In Fig. \ref{fig:3} we
plotted that component with respect to the {\em total momentum} of the
neutron obtained from both the 5.9 and 7.5 GeV/c beam momenta. 
The reason for choosing the total momentum was that the neutron momentum
is well measured and clearly represents the nucleon momentum distribution.
Each data point represents a single measured event.
The resolution of the vertical momentum component depends on the 
azimuthal angle of the pp scattering plane. The resolution is best if that
plane is horizontal and worst 
if it is vertical.  The different  sizes of the error bars in  Fig. 3
reflect this variation in the resolution.

The location of the neutron counter array below the targets insures that
the downward components of the neutron  momenta are nearly as large ($\sim$ 90\%)
as the total momenta.

   Any proton with a vertical component in the downward direction cannot be a
partner in the two-nucleon $np$ correlation. We see that, up to a neutron
momentum of $k_F \simeq 220$ MeV/c, (see the vertical dashed lines in Fig. 3)
 there are proton components in the downward as
well as in the upward directions. Above that momentum there are very few
downward pointing proton momenta and most point upwards.  This is what one
expects for  correlated nucleon pairs. The large neutron  momenta are
associated with upward going protons while below the Fermi level, where the
momenta can originate from the mean field, there is no preference. 
In order for neutrons to reach the counters below the targets they need
some minimal downward component $p_n^{vert}$, which increases with the total
neutron momentum. If the neutron is correlated with a proton, that proton
will need an upward component which is equal in size to  $p_n^{vert}$. The
slanted dashed line in the figure represents this transverse momentum
balance. All "correlated points "  would lie on that line if there were no
transverse motion of the $pn$ pair nor FSI. 
Correlated neutrons with a larger downward component
are associated with protons above that dashed line.

 The absence of downward pointing proton momenta at high
neutron momenta, is a clear indication of the dominance of two-nucleon
correlations.  We emphasize the fact that the conclusion is based only on
kinematics and does not depend on specific theoretical models. 

 We performed a series of tests to ascertain that the up/down asymmetry in
Fig. \ref{fig:3} is not an experimental artifact of the spectrometer or of
analysis procedures. We looked for asymmetries in, (i) the double
coincidence $(p,2p)$ quasi-elastic scattering data, (ii) the triple
coincidence data, where the third coincidence was taken from the photon
peak in the neutron counter and (iii) we changed the $E_{miss}$ region to
1.2 $< E_{miss} <$ 2 GeV.  We did not observe  asymmetries in any 
one, or combinations, of these tests.

 Final state interactions could, in principle, mimic this asymmetry. This
can happen if one of the outgoing protons scatters elastically from a
neutron in the same nucleus, at an angle such that the recoil neutron
enters the neutron counters. The momentum transferred to the proton cannot
be distinguished from the original momentum of the struck proton before 
the hard interaction.  Since the
neutron detectors are positioned at a backward angle (about $114 \pm 12$
deg), the probability for such a recoil neutron to enter the counters  is
very small. We estimated  the FSI  contribution to the  events shown in
Fig \ref{fig:3} in the following way. We assumed that the FSI can be
described by a Glauber-like calculation, as in Ref. \cite{kn:ref8}. We
simulated the geometry of the spectrometer and neutron detectors and
determined the number of neutrons and their momenta that could contribute
to Fig. \ref{fig:3}, for all the $(p,2p)$ quasi-elastic events accepted
by the spectrometer.
The  contribution of FSI to the events in Fig. 3 is about 3 above 220 MeV/c and 
about 2 below 220 MeV/c.
The number of  initial  state interactions for the 6 GeV/c projectile,
causing a neutron to recoil to an angle of 100 deg, is negligible.

 In conclusion, even with limited statistics,  it is possible to identify
two-nucleon short range correlations on an event by event basis.
 The identification is based
solely on kinematical arguments. The measured events are associated  with
high momentum components of the nuclear wave function.  It seems that  high
energy exclusive reactions are an effective tool for the study of SRC in
nuclei and it should encourage further studies, such as high energy $(e,epN)$
and $(p,2p+n)$ measurements.

Very special thanks are due to Dr. M. Sargsyan, who accompanied our
analysis  with detailed calculations.  We wish to thank Drs. L. Frankfurt
and M. Strikman for their theoretical input. S. Baker, 
F.J. Barbosa,
S. Kaye, M. Kmit, D. Martel, D. Maxam, J.E. Passaneau and M. Zilca  
contributed significantly to the design and construction of the detector.

We are pleased to acknowledge the assistance of the AGS staff
in building the detector and supporting the experiment, particularily our
liason engineers, J. Mills, D. Dayton, C. Pearson. We acknowledge the
continuing support of D. Lowenstein and P. Pile.

This research was supported by the U.S. - Israel Binational 
Science Foundation, the Israel Science Foundation founded by the 
Israel Academy of Sciences and Humanities, the NSF grant PHY-9501114 
and the U.S. Department of Energy grants DEFG0290  ER40553.

\end{document}